\documentclass[12pt,preprint]{aastex}

\shorttitle{Non-thermal Emissions from Cool Cores}
\shortauthors{Fujita \& Ohira}

\begin{document}

\title{Non-thermal Emissions from Cool Cores Heated by Cosmic-Rays in
Galaxy Clusters}

\author{Yutaka Fujita}
\affil{Department of Earth and Space Science, Graduate School of
Science, Osaka University, 1-1 Machikaneyama-cho, Toyonaka, Osaka
560-0043, Japan}
\email{fujita@vega.ess.sci.osaka-u.ac.jp}

\and

\author{Yutaka Ohira}
\affil{Theory Centre, Institute of Particle and Nuclear 
Studies, KEK, 1-1 Oho, Tsukuba 305-0801, Japan}

\begin{abstract}
 We study non-thermal emissions from cool cores in galaxy clusters. We
 adopted a recent model, in which cosmic-rays (CRs) prevail in the cores
 and stably heat them through CR streaming. The non-thermal emissions
 come from the interaction between CR protons and intracluster medium
 (ICM). Comparison between the theoretical predictions and radio
 observations shows that the overall CR spectra must be steep, and most
 of the CRs in the cores are low-energy CRs. Assuming that the CRs are
 injected through AGN activities, we study the nature of the shocks that
 are responsible for the CR acceleration. The steep CR spectra are
 likely to reflect the fact that the shocks travel in hot ICM with
 fairly small Much numbers. We also study the dependence on the CR
 streaming velocity. The results indicate that synchrotron emissions
 from secondary electrons should be observed as radio mini-halos in the
 cores. In particular, low-frequency observations (e.g. {\it LOFAR}) are
 promising. On the other hand, the steepness of the spectra makes it
 difficult to detect non-thermal X-ray and gamma-ray emissions from the
 cores. The low-energy CRs may be heating optical filaments observed in
 the cores.
\end{abstract}


\section{Introduction}

Clusters of galaxies are filled with hot X-ray gas or intracluster
medium (ICM) with temperatures of $\sim 2$--10~keV. While the radiative
cooling time of the ICM is longer than the age of the Universe in most
of the region in a cluster, the exception is the core, which is
$r\lesssim 100$~kpc from the cluster center. If there is no heating
source, the ICM in the core cools and a flow toward the cluster center
should develop (a cooling flow). However, X-ray observations have denied
the existence of massive cooling flows in clusters, which suggests that
the cores should be heated by some unknown sources
\citep[e.g.][]{ike97,mak01,pet01,tam01,kaa01,mat02}. Since active
galactic nuclei (AGNs) are often found in the cores, they are often
thought to be the heating sources
\citep*[e.g.][]{chu01,qui01,bru02,bas03}. X-ray observations have
actually revealed the interaction between AGNs and the ambient ICM
\citep*[e.g.][]{fab00,mcn00,bla01,mcn01,maz02,fuj02,joh02,kem02,tak03,fuj04}. However,
even if AGNs can produce enough energy to heat the core, the energy must
deliberately be transported to the surrounding ICM in the core. For
example, conventional mechanical heating such as the dissipation of weak
shocks and sound waves often cause thermal instabilities
\citep*{fuj05,mat06}. Therefore, turbulence may essentially be required
to hold the instabilities for such heating mechanisms.

Cosmic-rays (CRs) may be another channel of transporting energy to the
ICM \citep*[e.g.][]{tuc83,rep87,rep95,col04,pfr07,jub08}. Especially, CR
streaming has been studied as an energy transport mechanism
\citep*{rep79,boh88,loe91,guo08}. In this mechanism, CRs streaming in
the ICM excites Alfv\'en waves. The CRs interact and move outwards with
the waves. The $PdV$ work done by the CRs effectively heat the
ICM. Recently, using numerical simulations, we showed that the CR
streaming can stably heat the core for a long time \citep[][hereafter
Paper~I]{fuj11}. The reason of the stability is that the CR pressure is
insensitive to changes in the ICM and that the density dependence of the
heating term is similar to that of radiative cooling. Moreover, CRs can
prevail in the entire core and the heating is not localized around the
source. The CRs may be provided in the core not only by AGNs but also
through pumping by turbulence \citep{ens11}.

In this paper, we study the non-thermal emission from the CRs that heat
cool cores and the AGN activities that are responsible for the
acceleration of the CRs. It is to be noted that non-thermal emissions
from CR protons accelerated by AGNs in the cores have been studied by
\citet{fuj07c}. However, they studied CR acceleration associated with a
single AGN burst with an extremely large energy, and they did not
consider the heating of the ICM by CR streaming. This paper is organized
as follows. In \S~\ref{sec:model}, we explain our models on core heating
and AGN activities that are responsible for the generation of CRs. In
\S~\ref{sec:AGNr}, we present the results of our calculations and
compare them with observations. In \S~\ref{sec:dis}, we discuss the
implications of our results, and \S~\ref{sec:conc} is devoted to
conclusions. We refer to protons as CRs unless otherwise mentioned.

\section{Models}
\label{sec:model}

\subsection{Cosmic-Ray Distributions}
\label{sec:CRdist}

In Paper~I, we studied heating of a cool core by CRs injected through
the activities of the central AGN. The CRs travel with Alfv\'en waves in
the ICM. They amplify the waves, which heat the surrounding ICM. In this
subsection, we briefly summarize the models to obtain CR and ICM
distributions.

For simplicity, we assumed that the cluster is spherically symmetric.
The flow equations are
\begin{equation}
 \frac{\partial \rho}{\partial t} 
+ \frac{1}{r^2}\frac{\partial}{\partial r}(r^2\rho u) = 0\:,
\end{equation}
\begin{equation}
\frac{\partial (\rho u)}{\partial t} 
+ \frac{1}{r^2}\frac{\partial}{\partial r}(r^2\rho u^2)
= - \rho \frac{G M(r)}{r^2}-\frac{\partial}{\partial r}
(P_g + P_c + P_B)\:,
\end{equation}
\begin{eqnarray}
 \frac{\partial e_g}{\partial t}  
+ \frac{1}{r^2}\frac{\partial}{\partial r}(r^2 u e_g)
&=& -P_g \frac{1}{r^2}\frac{\partial}{\partial r}(r^2 u) 
+ \frac{1}{r^2}\frac{\partial}{\partial r}
\left[r^2\kappa(T)\frac{\partial T}{\partial r}\right]\nonumber\\
& &- n_e^2\Lambda(T) + H_{\rm st} + H_{\rm coll}\:,
\label{eq:eg}
\end{eqnarray}
\begin{equation}
\label{eq:ec}
 \frac{\partial e_c}{\partial t}  
+ \frac{1}{r^2}\frac{\partial}{\partial r}(r^2 \tilde{u} e_c)
= -P_c \frac{1}{r^2}\frac{\partial}{\partial r}(r^2 \tilde{u}) 
+ \frac{1}{r^2}\frac{\partial}{\partial r}
\left[r^2 D(\rho)\frac{\partial e_c}{\partial r}\right] 
- \Gamma_{\rm loss}
+ \dot{S}_c \:,
\end{equation}
where $\rho$ is the gas density, $u$ is the gas velocity, $P_g$ is the
gas pressure, $P_c$ is the CR pressure, $P_B$ is the magnetic pressure,
$G$ is the gravitational constant, $M(r)$ is the gravitational mass
within the radius $r$, $\kappa(T)$ is the coefficient for thermal
conduction and $T$ is the temperature, $n_e$ is the electron density,
$\Lambda$ is the cooling function, $H_{\rm st}$ is the heating by CR
streaming, $H_{\rm coll}$ is the heating by Coulomb and hadronic
collisions, $\tilde{u}$ is the CR transport velocity, $D(\rho)$ is the
diffusion coefficient for CRs averaged over the CR spectrum,
$\Gamma_{\rm loss}$ is the energy loss by Coulomb and hadronic
collisions, and $\dot{S}_c$ is the source term of CRs. Energy densities
of the gas and the CRs are respectively defined as
$e_g=P_g/(\gamma_g-1)$ and $e_c=P_c/(\gamma_c-1)$, where $\gamma_g=5/3$
and $\gamma_c=4/3$. In this paper, we do not treat models with thermal
conduction, and thus $\kappa=0$. The terms for radiative cooling
$\Lambda$, Coulomb collisions $H_{\rm coll}$, hadronic collisions
$H_{\rm coll}$, diffusion $D(P)$, and the energy loss $\Gamma_{\rm
loss}$ are the same as those in Paper~I. The source term of CRs is given
by $\dot{S}_c \propto L_{\rm AGN}$, where $L_{\rm AGN}$ is the energy
injection rate from the AGN. We assume that $L_{\rm AGN}=\epsilon
\dot{M} c^2$, where $\epsilon$ is the parameter, $\dot{M}$ is the inflow
rate of the gas toward the AGN, and $c$ is the speed of light.

The CR transport velocity in equation~(\ref{eq:ec}) is given by
$\tilde{u}=u+v_A$, where $v_A=B/\sqrt{4\pi\rho}$ is the Alfv\'en
velocity for a magnetic field $B$, which evolves as $B\propto
\rho^{2/3}$. The initial magnetic field at the cluster center is
$B_0=10\: \mu G$. The wave energy $U_A=\delta B^2/(4\pi)$, where $\delta
B$ is the magnetic field fluctuation, is amplified by the $PdV$ work
done by the CRs on Alfv\'en waves:
\begin{equation}
\label{eq:UA}
 \frac{\partial U_A}{\partial t} + \frac{1}{r^2}
\frac{\partial}{\partial r}
\left[r^2 U_A\left(\frac{3}{2}u + v_A\right)\right]
= u\frac{\partial}{\partial r}\frac{U_A}{2} -
v_A\frac{\partial P_c}{\partial r} - H_{\rm st} \:
\end{equation}
\citep{mck82,boh88}. This equation is more correct than that we adopted
in Paper~I (equation~[6] in that paper), because it is based on wave
energy conservation. However, the results are not affected by this
change of the equation (see \S~\ref{sec:AGNr}). After the wave energy
increases to $U_A\sim U_M$, where $U_M$ is the energy of the background
magnetic field, the waves are expected to heat ICM through non-linear
damping \citep[e.g.][]{ohira09,gargate10}. Thus, we give the heating
term for CR streaming by
\begin{equation}
 H_{\rm st} = \Gamma v_A\left|\frac{\partial P_c}{\partial r}\right|
\end{equation}
\citep{voe84,kan06}. We simply give $\Gamma=U_A/U_M$ for $U_A<U_M$ and
$\Gamma=1$ after $U_A$ reaches $U_M$.

\subsection{Non-Thermal Emissions}
\label{sec:AGNm}

Although by solving equations presented in \S~\ref{sec:CRdist} we can
obtain the profile of the ICM and that of the CR pressure $P_c(r)$
required to heat the core effectively (Paper~I), we do not have
information on the energy spectrum of the CRs. Thus, we need to specify
the spectrum of the CRs to calculate the non-thermal emissions from the
CRs.

We assume that the central AGN drives outgoing shock waves and form
cocoons or bubbles inside them. In the following, we show a description
of their evolution (position, velocity, Mach number as functions of
time). The shocks should inject CRs with varying efficiencies and
spectra. At some moment this injection will be maximal (actually, this
moment differs for different CR energy ranges). We only consider the
efficiency and Mach number at this moment and fix these numbers by
requesting them to reproduce the observed radio emission for the sake of
simplicity, although there would be a more physical approach to
calculate the injection evolution and the full injected CR spectrum,
assuming the AGN energy release, timescale, and initial cocoon radius.

The CRs are accumulated in the core through the AGN activities. In
Paper~I, we studied continuous CR injection as a time average, although
the supply of the CRs may be intermittent. Each activity of the AGN is
approximated by an instantaneous explosion. Thus, the shock expands in
the ICM like a supernova remnant in the Galaxy, and the shock velocity
depends on the energy input from the AGN.

In Paper~I, we obtained the profiles of the ICM density $\rho(r)$, the
temperature $T(r)$, and the magnetic field $B(r)$ at a given time. We
approximate the density profile of the ICM by a power-law:
\begin{equation}
 \rho_{\rm ICM}(r)=\rho_{\rm in}(r/r_{\rm in})^{-\omega}\:,
\end{equation}
where $\rho_{\rm in}$ is the ICM density at the inner boundary ($r_{\rm
in}=5$~kpc). Using a shell approximation \citep[e.g.][]{ost88}, the
radius of the shock can be written as
\begin{equation}
\label{eq:Rs}
 R_s = \xi\left(\frac{E_a}{\rho_{\rm in} r_{\rm in}^\omega}\right)
^{1/(5-\omega)} t_a^{2/(5-\omega)}\;,
\end{equation}
where
\begin{equation}
 \xi = \left[\left(\frac{5-\omega}{2}\right)^2\frac{3}{4\pi}
\frac{(\gamma_g+1)^2(\gamma_g-1)(3-\omega)}{9\gamma_g-3-\omega(\gamma_g+1)}
\right]^{1/(5-\omega)}\:,
\end{equation}
$E_a$ is the energy released by the AGN, and $t_a$ is the time elapsed
since the last energy input from the AGN. The velocity of the shock is
given by
\begin{equation}
\label{eq:Vs}
 V_s = \frac{d R_s}{d t_a} 
\:.
\end{equation}
The Mach number of the shock is given by $M_s=V_s/c_s(R_s)$, where $c_s$
is the sound velocity. Since we know the profile of the ICM temperature
$T(r)$, we can construct the profile of the sound velocity
$c_s(r)$. Therefore, if $M_s$ and $E_a$ are given, the shock radius
$R_s$, velocity $V_s$, and the time $t_a$ that satisfy
equations~(\ref{eq:Rs}) and~(\ref{eq:Vs}) can be specified.

In reality, the spectrum of accelerated CRs at the shock may change
during the expansion of the cocoon. Probably, the spectrum is flat, when
the cocoon is young, and the shock velocity and the Mach number are
large. Then it gradually steepens as the Mach number decreases, and CR
acceleration ceases when the Mach number approaches $M_s\sim
1$. However, we consider a typical Mach number $M_{st}$ around which
most CRs are accelerated. In other words, we consider a typical spectrum
of CRs that are accelerated when the injection of CRs becomes
maximal. We treat $M_{st}$ and $E_a$ as parameters. The shock radius,
velocity, and age when $M_s=M_{st}$ are $R_s=R_{st}$, $V_s=V_{st}$, and
$t_a=t_{at}$, respectively. We also assume that the spectrum of CRs that
are just accelerated at $r\sim R_{st}$ has a form of
\begin{equation}
\label{eq:NE}
 N(p,r)\propto p^{-x}e^{-p/p_{\rm max}}\:,
\end{equation}
where $p$ is the CR momentum, $x$ is the index, and $p_{\rm max}$ is the
cutoff momentum of the CRs. Since we already know CR pressure $P_c(r)$,
the normalization of relation~(\ref{eq:NE}) is determined by the
relation
\begin{equation}
\label{eq:Pc}
 P_c(r) = \frac{c}{3}\int_{p_{\rm min}}^\infty
\frac{p^2 N(p,r)}{\sqrt{p^2 + m^2 c^2}} dp\:,
\end{equation}
where $m$ is the proton mass. 

The index is given by $x=(r_b+2)/(r_b-1)$, where
$r_b$ is the compression ratio of the shock \citep{bla87}, which is
given by
\begin{equation}
\label{eq:comp}
 r_b = \frac{(\gamma_g+1) M_{st}^2}{(\gamma_g-1) M_{st}^2+1}\:.
\end{equation}
Since the cooling of CR protons is not effective, the maximum energy of
protons corresponding to $p_{\rm max}$ is determined by the age of the
shock and is represented by
\begin{equation}
 E_{\rm max}\sim 1.6\times 10^4\: 
\left(\frac{V_{st}}{10^3\rm\: km\: s^{-1}}\right)^2
\left(\frac{B_d}{10\: \mu\rm\: G}\right)
\left(\frac{t_{at}}{10^7\:\rm yr}\right)\:{\rm TeV}\:,
\label{eq:pmax}
\end{equation}
where $B_d$ is the downstream magnetic field at $r=R_{st}$ and is given
by $B_d = r_b B$ \citep{yam06,fuj07c}. That is, the background magnetic
field $B$ is amplified by the compression ratio $r_b$
(equation~[\ref{eq:comp}]). Although some particles may be accelerated
to higher energies when the expansion velocity of the cocoon was larger,
their contribution to the overall spectrum is expected to be small.

The CRs injected at $r\sim R_{st}$ propagate in the ICM with Alfv\'en
waves. Although adiabatic cooling may change $E_{\rm max}$, it does not
change the index $x$ in relation~(\ref{eq:NE}). Moreover, the results in
\S~\ref{sec:AGNr} show that the CR spectra must be steep. Thus, the
results are not sensitive to the value of $E_{\rm max}$. Therefore, we
do not consider the adiabatic cooling for $E_{\rm max}$ and adopt the
relations~(\ref{eq:NE}) and~(\ref{eq:Pc}) at any radius $r$, although
the adiabatic cooling was considered when we calculated $P_c$ in
Paper~I. Since we expect that thermal protons with higher energies are
accelerated as CRs, we assume that the minimum momentum of the CRs is
$p_{\rm min}=4 m c_{sd}$, where $c_{sd}$ is the sound velocity of the
ICM at the downstream of the shock at $r=R_{st}$, which is obtained from
the Rankine-Hugoniot relations for given $c_s(R_{st})$ and $M_{st}$:
\begin{equation}
 c_{sd} = c_s\frac{\sqrt{2\gamma_g M_{st}^2 
- (\gamma_g-1)}\sqrt{(\gamma_g-1)
  M_{st}^2 + 2}}{(\gamma_g + 1)M_{st}}\:.
\end{equation}
In this way we have the CR spectrum at each radius for given $M_{st}$
and $E_a$. 

For a given CR proton spectrum, we calculate radiation from
them. Non-thermal emission originated from CR protons in the central
region of clusters have been studied by several groups
\citep[e.g.][]{min03,pfr04,kes10}. In this paper, we adopt the model of
\citet{fuj09c}, in which they calculated non-thermal emissions from
supernova remnants. We consider the synchrotron, bremsstrahlung, and
inverse Compton (IC) emissions from secondary electrons created through
the decay of charged pions that are generated through proton-proton
collisions. IC emissions are created by electrons that scatter Cosmic
Microwave Background (CMB) photons. We also consider $\pi^0$-decay
gamma-rays through proton-proton collisions. We do not consider
emissions from primary electrons that are directly accelerated at the
shock, because we did not calculate the distribution of the primary
electrons in Paper~I. Because of the short cooling time of electrons,
emissions from primary electrons will disappear soon after their
acceleration is finished \citep{fuj07c}.

The photon spectra are calculated based on the radiation models of
\citet{fan07}. For the production of secondary electrons and
$\pi^0$-decay gamma-ray photons through proton-proton interactions, we
use the code provided by \citet{kar08}. The spectrum of the secondary
electrons are given by $N_e(E_e)=t_{\rm cool,e}(E_e)\: Q_e(E_e)$, where
$E_e$ is the electron energy, $t_{\rm cool,e}$ is the cooling time of an
electron, and $Q_e$ is the production rate of the secondary electrons.
For the cooling, we include synchrotron cooling, IC scattering,
Bremsstrahlung, and Coulomb loss.

\section{Results}
\label{sec:AGNr}

\begin{figure}
\epsscale{.80}
\plotone{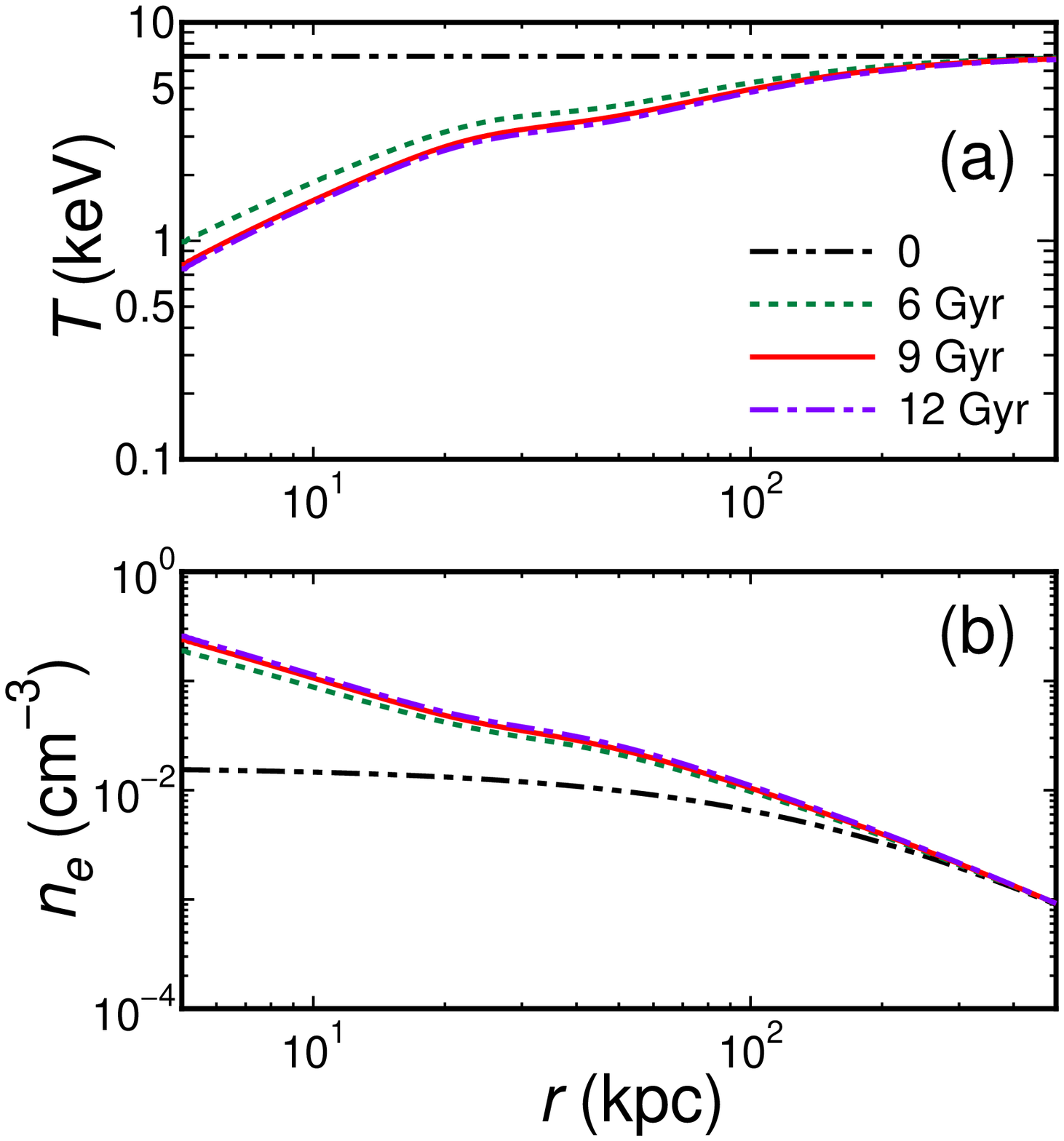}
\caption{(a) Temperature and (b) density profiles for Model~LCR0.}
\label{fig:Tn_lcr0}
\end{figure}

\begin{figure}
\epsscale{.80} 
\plotone{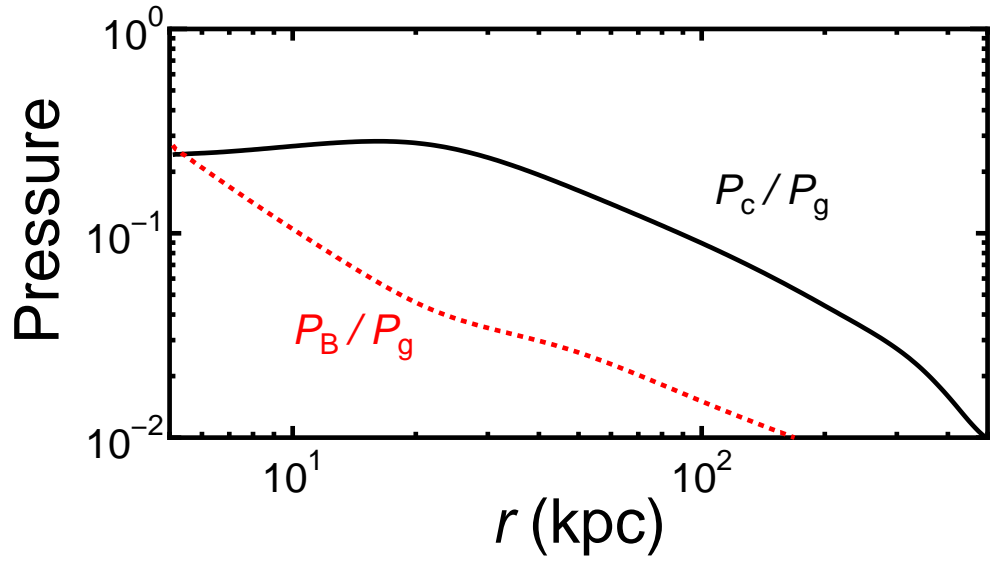} 
\caption{Profiles of the ratios
$P_c/P_g$ (solid) and $P_B/P_g$ (dotted) at $t=9$~Gyr for Model~LCR0.}
\label{fig:Pcb_lcr0}
\end{figure}

Since we replaced the equation for the wave energy $U_A$ (equation~[6]
in Paper~I) with equation~(\ref{eq:UA}), we recalculate the
distributions of the ICM and CRs and show them in
Figures~\ref{fig:Tn_lcr0} and~\ref{fig:Pcb_lcr0}. The cluster is
initially isothermal with $P_c=0$. The input parameters are the same as
those of Model~LCR0 in Paper~I, and we simply refer to this model as
LCR0 again. The gravitational potential adopted in this model is that of
the Perseus cluster. The efficiency of AGN energy input is
$\epsilon=2.5\times 10^{-4}$. The results are almost identical to those
in Paper~I (Figures~2 and 4 in that paper). This is because $\Gamma$
rapidly approaches one after $L_{\rm AGN}$ increases regardless of the
equation we adopted. The ICM temperature outside the core is $\sim
7$~keV. The heating by CR streaming and the radiative cooling are
well-balanced at $t\gtrsim 4$~Gyr.

Figure~\ref{fig:lcr0} shows the spectra of a region including the entire
core ($r<1$~Mpc) at $t=9$~Gyr for  Model LCR0. The slope of the ICM
density profile is assumed to be $\omega=1$, which is a good
approximation for $r\lesssim 70$~kpc (Figure~\ref{fig:Tn_lcr0}b). The
distance to the cluster is 78.4~Mpc, which is the one for the Perseus
cluster. In this figure, we take $M_{st}=2.1$ and $E_a=1\times
10^{60}\rm\: erg\: s^{-1}$; we first give $E_a$, and then adjust
$M_{st}$ in order to be consistent with radio observations for the
mini-halo in the Perseus cluster \citep*{sij93,git02}. Since the Mach
umber $M_{st}$ is fairly small, the CR spectrum is steep ($x=3.2$). The
maximum energy of the CRs is $E_{\rm max}=1.5\times 10^5$~TeV, the
radius and age of the shock are $R_{st}=22$~kpc and $t_{at}=6.0\times
10^6$~yr, respectively. The spectrum of thermal Bremsstrahlung is shown
for comparison.

The slope of the synchrotron and IC scattering spectra at the higher
energy side can be explained by a simple calculation. The slope of the
energy spectrum of secondary electrons are the same as that of protons
($x=3.2$), if radiative cooling is not effective. However, cooling by
synchrotron radiation and IC scattering increases the slope by one and
it becomes $x'=4.2$ \citep[e.g.][]{sar99}. The spectral indices of the
synchrotron emission and IC scattering are represented by
$\alpha=(x'-1)/2=1.6$ \citep[e.g.][]{ryb79}, which is consistent with
those in Figure~\ref{fig:lcr0} ($f_\nu\propto \nu^{-\alpha}$).

We found that the results for $M_{st}=2.1$ and $E_a=1\times 10^{61}\rm\:
erg\: s^{-1}$ are almost the same as those for $M_{st}=2.1$ and
$E_a=1\times 10^{60}\rm\: erg\: s^{-1}$. In the former case, the shock
radius and age are $R_{st}=59$~kpc and $t_{at}=1.4\times 10^7$~yr,
respectively. Although the maximum energy of the CRs is increased
($E_{\rm max}=2.7\times 10^5$~TeV), the steep CR spectrum or the large
$x$ obscures the effect. This means that the radiation from the cool
core is insensitive to the strength of an AGN activity ($E_a$) for a
{\it given} $P_c(r)$. Figure~\ref{fig:lcr0_2} shows the spectra when
$M_s=1.8$ and 4.0 for $E_a=1\times 10^{60}\rm\: erg\: s^{-1}$. The
indices are $x=3.8$ and $2.3$, respectively. Compared with the results
of $M_s=2.1$ and $E_a=1\times 10^{60}\rm\: erg\: s^{-1}$, the
non-thermal emissions are weaker (stronger) when $M_s$ is smaller
(larger), and the synchrotron radio emission is inconsistent with the
observations. The results are very sensitive to the value of $M_{st}$
for a given $P_c(r)$. Basically, changing $E_a$ and $M_{st}$ correspond
to changing $E_{\rm max}$ and $x$, respectively.

Figure~\ref{fig:surf_lcr0} shows the surface brightness profiles for
Model~LCR0 at $t=9$~Gyr with $M_{st}=2.1$ and $E_a=1\times 10^{60}\rm\:
erg\: s^{-1}$. The model generally reproduces the surface brightness
profile observed in the radio band, although we did not intend to
reproduce that when we calculated Model~LCR0 in Paper~I. In that figure,
the surface brightness rapidly increases toward the cluster center for
the synchrotron radio emissions because of the increase of the magnetic
fields toward the cluster center ($B(r)\propto\rho(r)^{2/3}$). On the
other hand, the profile for the IC missions is relatively flat because
electrons scatter CMB photons, which are uniformly distributed. The size
of the region with high surface brightness is regulated by radiative
cooling, because radiative cooling increases the ICM density and makes a
cool core. On the other hand, CRs can fill the entire core with fast
Alfv\'en waves (Paper~I). Proton-proton interactions are effective in
such a high-density region. The surface brightness for thermal
Bremsstrahlung slightly decreases at the cluster center, because the ICM
temperature decreases there (Figure~\ref{fig:Tn_lcr0}a).

We also study a less massive cluster. Figure~\ref{fig:scr0} shows the
spectra of the entire core ($r<1$~Mpc) at $t=9$~Gyr calculated using
parameters of Model~SCR0 in Paper~I. For this model, we adopted the
observed gravitational potential of the Virgo cluster. The efficiency of
AGN energy input is $\epsilon=1\times 10^{-4}$. Although we
recalculated the ICM and CR distributions, they are almost identical to
those calculated in Paper~I. The ICM temperature outside the core is
$\sim 2$~keV. We take $M_{st}=2.1$ ($x=3.2$) and $E_a=1\times
10^{59}\rm\: erg\: s^{-1}$. The distance to the cluster is set to be
16~Mpc. We take $\omega=0.7$, which is an good approximation for
$r\lesssim 50$~kpc (Figure~11 in Paper~I). The shock radius and age are
$R_{st}=21$~kpc and $t_{at}=6.3\times 10^6$~yr, respectively. The
maximum energy of CRs is $E_{\rm max}=3.3\times 10^4$~TeV. The gamma-ray
flux is much smaller than the upper limits for the Virgo cluster
obtained with {\it Fermi} \citep{ack10}. The luminosity is sensitive to
$M_{st}$ but not to $E_a$. The surface brightness profiles for this
model are shown in Figure~\ref{fig:surf_scr0}. The surface brightness is
smaller than that in Figure~\ref{fig:surf_lcr0}.

\section{Discussion}
\label{sec:dis}

We have studied the non-thermal spectra of cool cores heated by CR
streaming. The results indicate that the Mach number of the shock that
accelerate CRs must be small ($\sim 2$) to be consistent with radio
observations at least for the Perseus cluster. We think that this is
reasonable because the temperature and the sound velocity of the ICM is
large and thus it is difficult for the cocoon shock to have a large Mach
number. The small Mach number means that the CR spectrum must be steep.
In Paper~I, we did not specify the injection mechanism of CRs. We
emphasize that even if CRs are injected by anything other than the
cocoon, the spectrum must be steep for the given $P_c$.

\begin{figure}
\epsscale{.80} \plotone{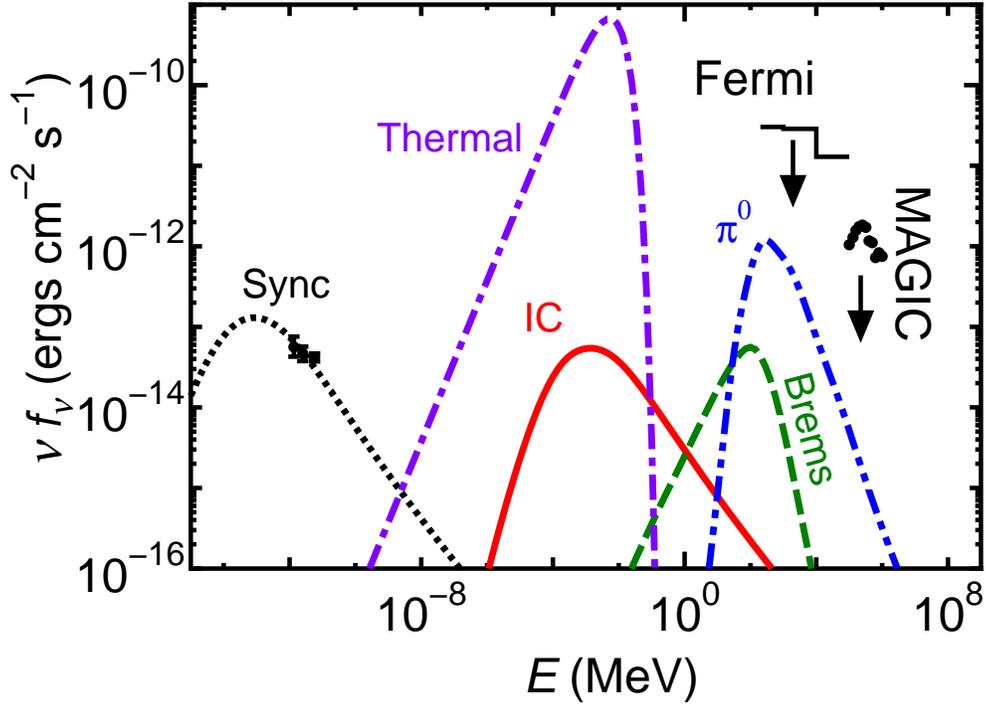} \caption{ Spectra calculated based on
Model~LCR0 with $M_{st}=2.1$ and $E_a=1\times 10^{60}\rm\: erg\:
s^{-1}$.  The synchrotron radiation (dotted line), IC scattering (solid
line) and non-thermal bremsstrahlung (dashed line) are of the secondary
electrons. The $\pi^0$ decay gamma-rays are shown by the two-dot-dashed
line. The thermal bremsstrahlung is shown by the dot-dashed
line. Observations are for the Perseus cluster. Radio observations are
shown by dots \citep{sij93,git02}, and gamma-ray upper limits are shown
by arrows \citep{ack10,ale10}.}  \label{fig:lcr0}
\end{figure}

\begin{figure}
\epsscale{1.0} \plottwo{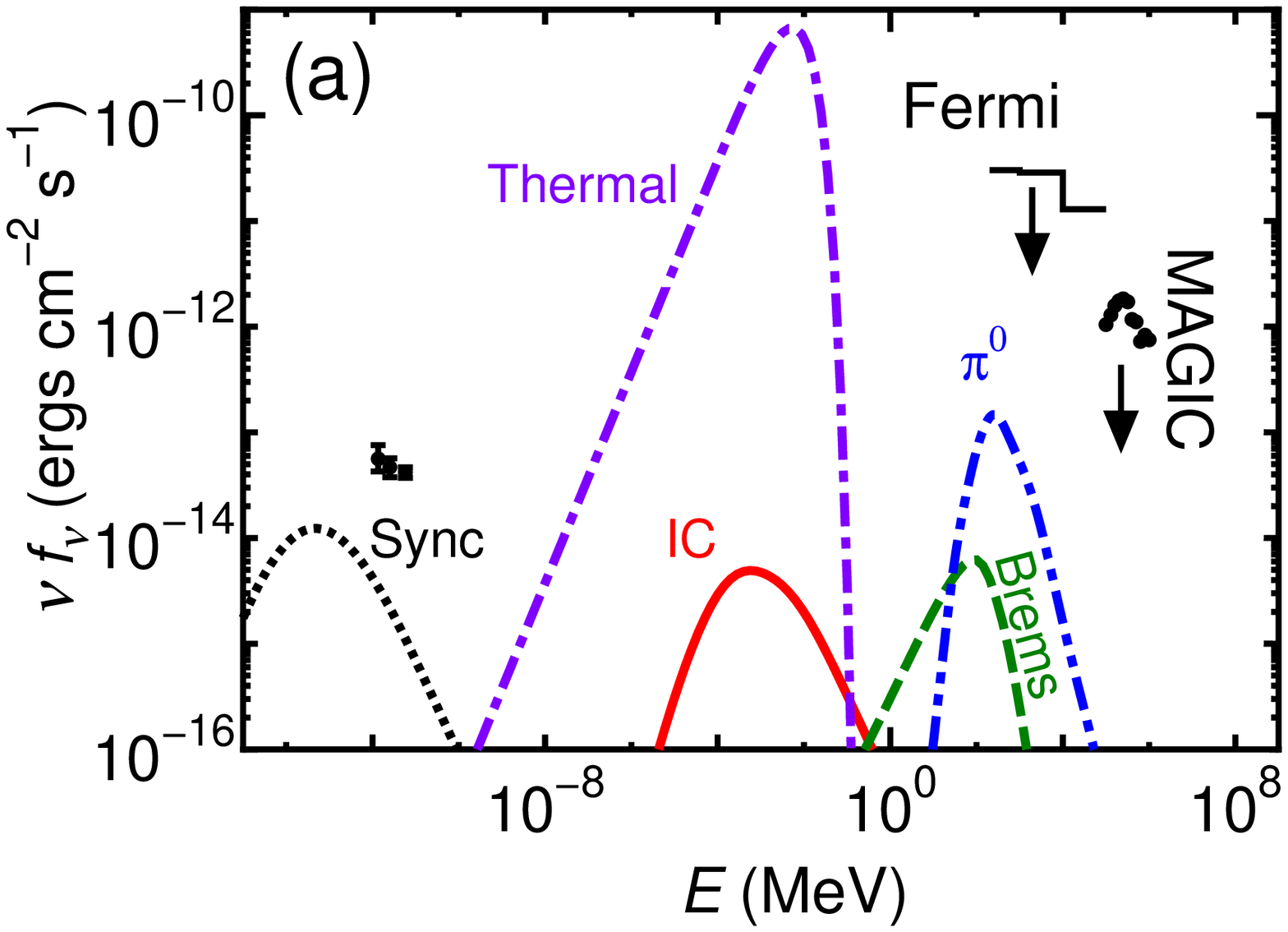}{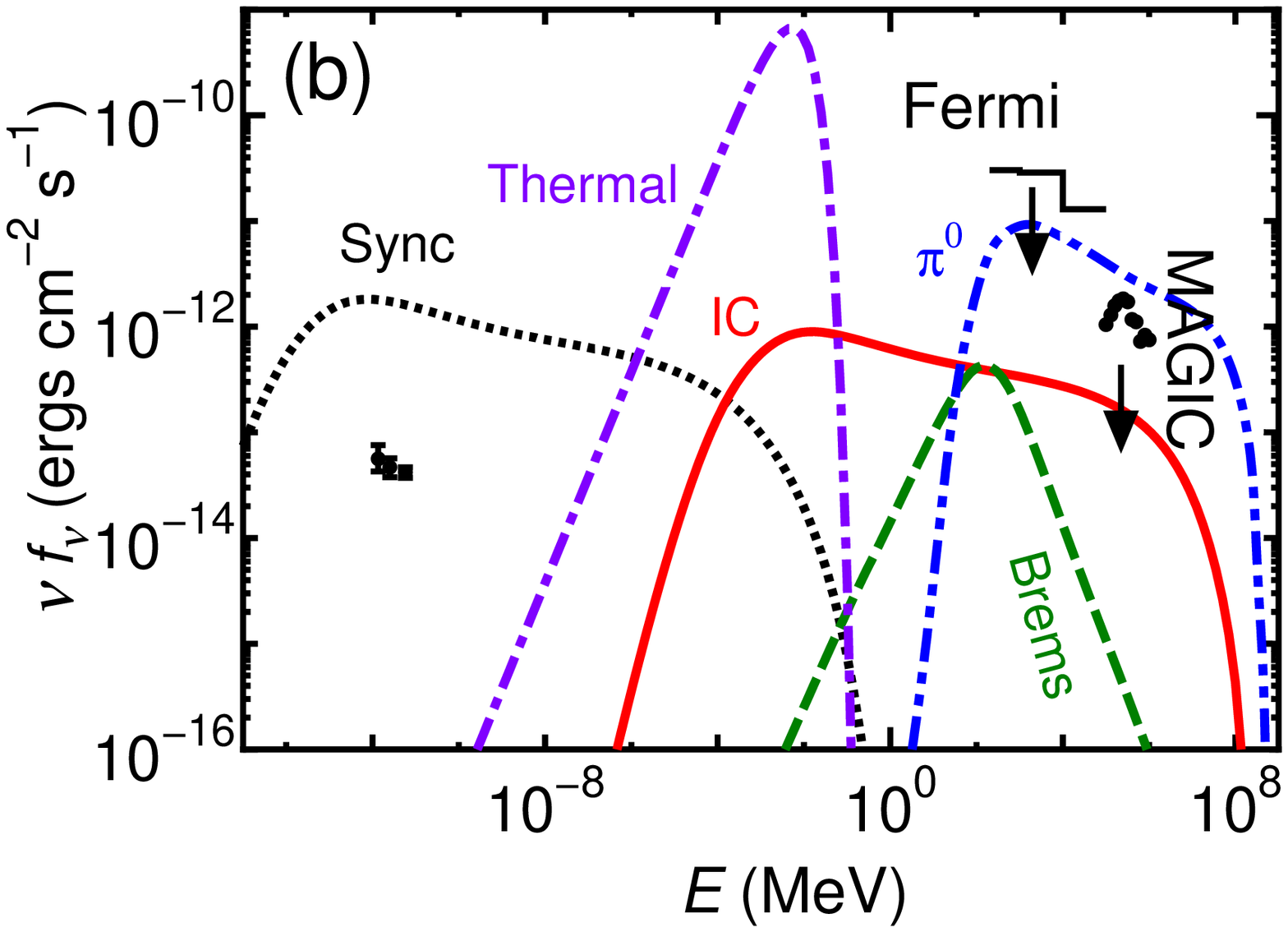} \caption{Same as
Fig.~\ref{fig:lcr0} but for (a) $M_s=1.8$ and (b) 4.0.}
\label{fig:lcr0_2}
\end{figure}

\begin{figure}
\epsscale{.80} \plotone{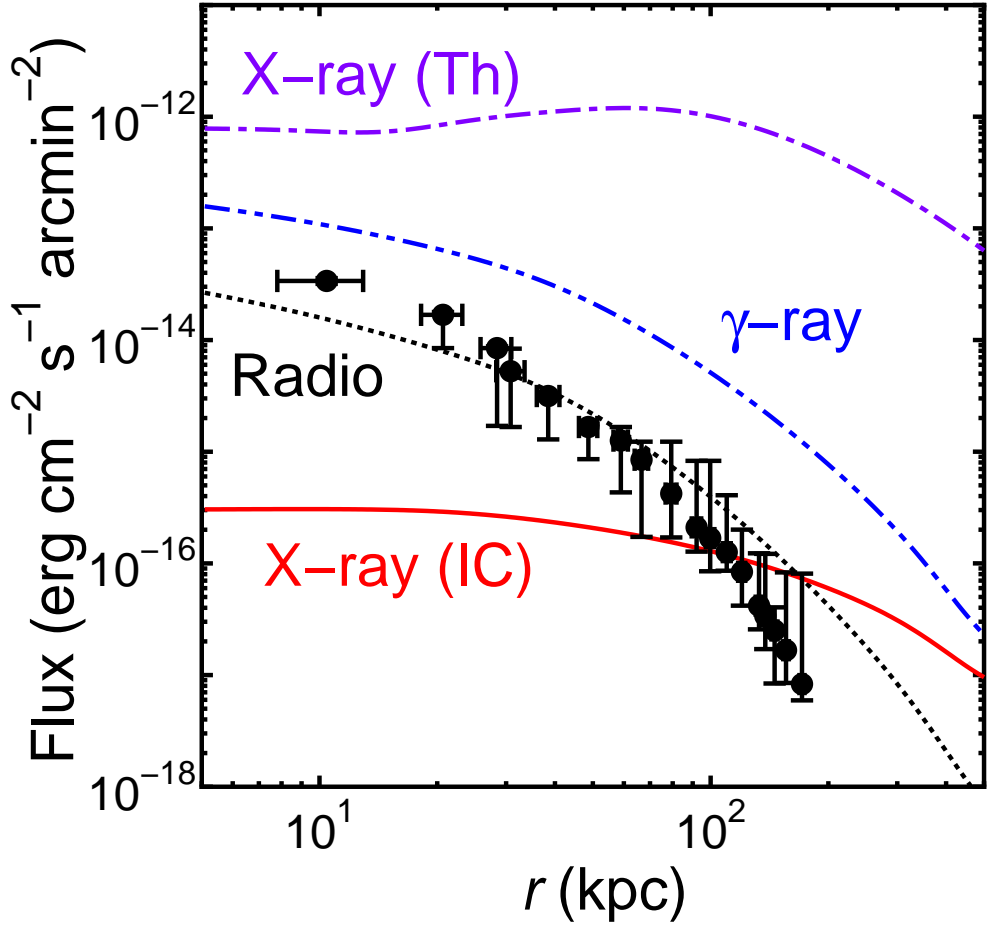} \caption{Surface brightness of
 non-thermal and thermal emissions calculated based on Model~LCR0 with
 $M_{st}=2.1$ and $E_a=1\times 10^{60}\rm\: erg\: s^{-1}$. The
 synchrotron radiation (327~MHz; dotted line), IC scattering (20~keV;
 solid line), $\pi^0$ decay gamma-rays (1~GeV; two-dot-dashed line), and
 thermal Bremsstrahlung (20~keV; dot-dashed line) are shown. Radio
 observations for the mini-halo in the Perseus cluster are shown by dots
 \citep*{git03}. The vertical errors include the deviations from the
 spherical symmetry.}  \label{fig:surf_lcr0}
\end{figure}

Recently, \citet{ens11} indicated that the CR streaming velocity may be
much larger than the Alfv\'en velocity $v_A$ in the hot ICM. This is
because in high-$\beta$ plasma, where $\beta$ is the ratio of thermal to
magnetic energy, waves may suffer strong resonant damping by thermal
protons. In this case, the sound velocity $c_s$ may be appropriate as
the streaming velocity instead of $v_A$ \citep{hol79,ens11}. Thus, we
simply replace $v_A$ by $c_s$ in equations~(\ref{eq:ec})
and~(\ref{eq:UA}) and see what would happen.  Figures~\ref{fig:Tn_lcrs}
and~\ref{fig:Pcb_lcrs} show the profiles of the ICM and CRs for the
parameters of Model~LCR0 except for the larger streaming velocity $c_s$;
we refer to this model as Model~LCRs. The ICM is stably heated by the CR
streaming even in this model and the evolution of $\dot{M}$ is not much
different from that in Model LCR0. Compared with
Figures~\ref{fig:Pcb_lcr0}, the fraction of CR pressure is small in the
central region because of the larger streaming velocity and the escape
of CRs. Since the ICM temperature is an increase function of radius
(Figure~\ref{fig:Tn_lcrs}), the sound velocity or the streaming velocity
is also an increasing function. Thus, $P_c/P_g$ tends to decrease
outward fairly rapidly. Figures~\ref{fig:lcrs} and~\ref{fig:surf_lcrs}
are the same as Figures~\ref{fig:lcr0} and~\ref{fig:surf_lcr0} but for
Model LCRs. If we assume $M_{st}=2.1$ as is the case of Model LCR0,
non-thermal luminosities in Model LCRs are smaller than those in Model
LCR0, because more CRs have escaped from the core with the high ICM
density. Thus, we increase the Mach number and set it to be
$M_{st}=2.4$.  We present the spectra and surface brightness in
Figures~\ref{fig:lcrs} and~\ref{fig:surf_lcrs}. The synchrotron
spectrum and the surface brightness are consistent with the
observations.

\begin{figure}
\epsscale{.80} \plotone{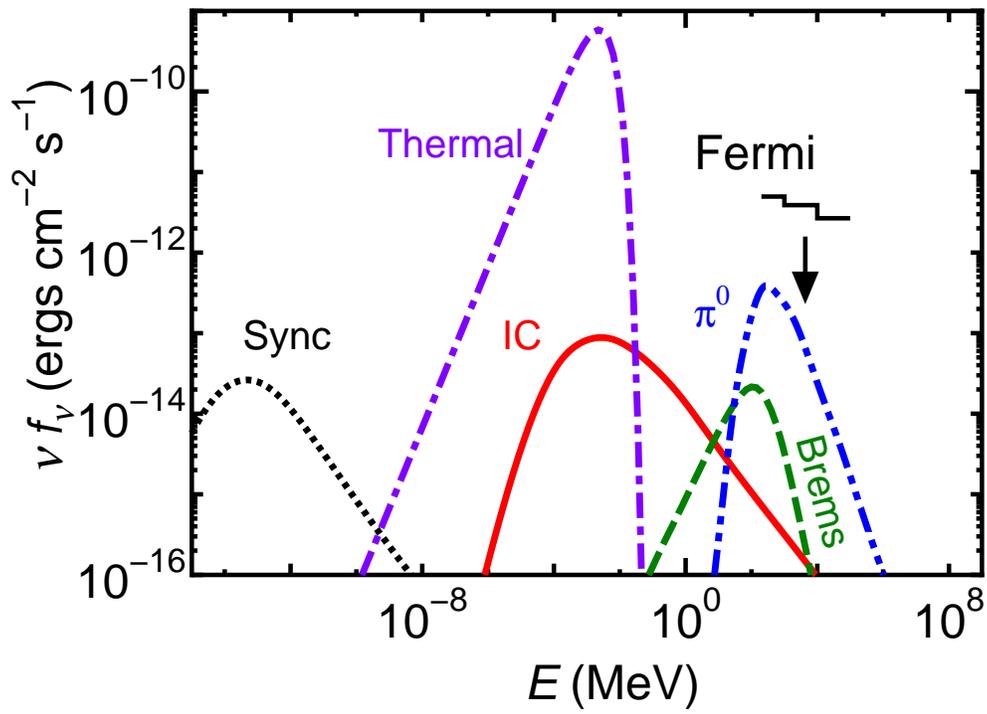} \caption{Same as Fig.~\ref{fig:lcr0} but
for Model~SCR0. Gamma-ray upper limits are shown by an arrow
\citep{ack10}.} \label{fig:scr0}
\end{figure}

\begin{figure}
\epsscale{.80} \plotone{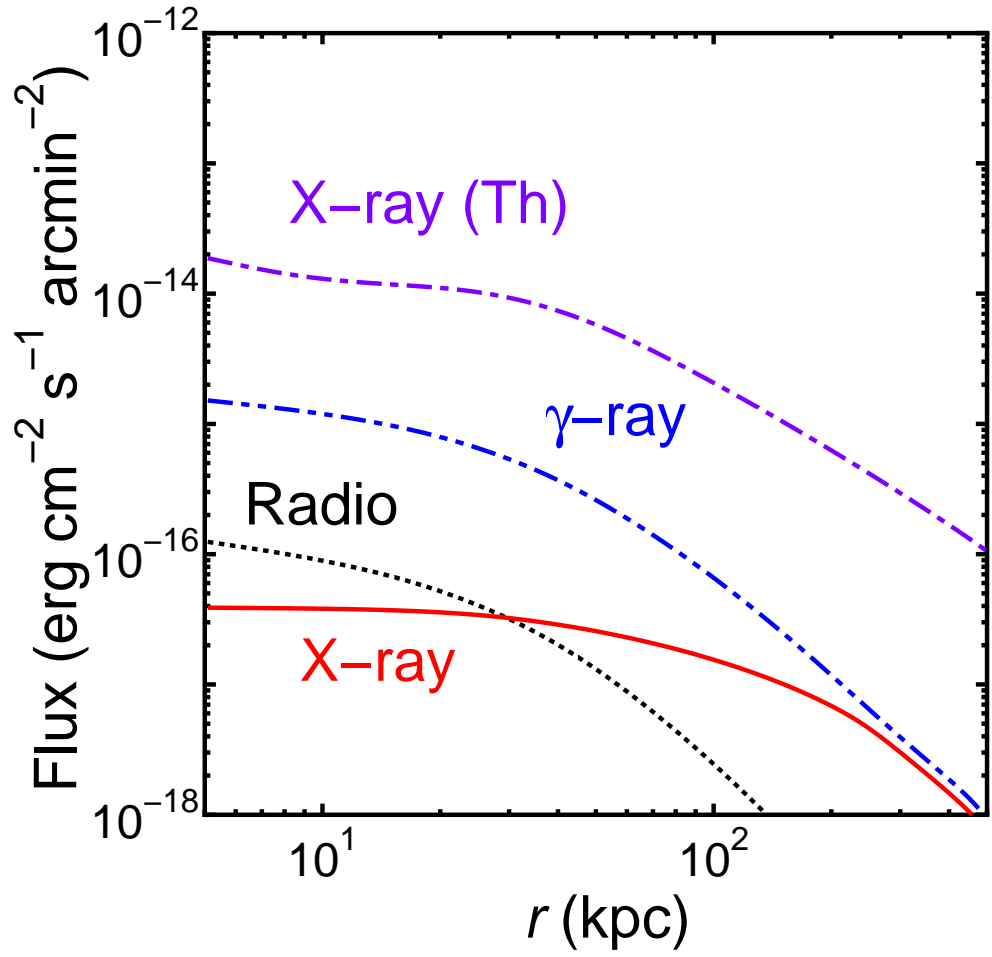} \caption{Same as Fig.~\ref{fig:scr0} but
for Model~SCR0.} \label{fig:surf_scr0}
\end{figure}

\begin{figure}
\epsscale{.80} \plotone{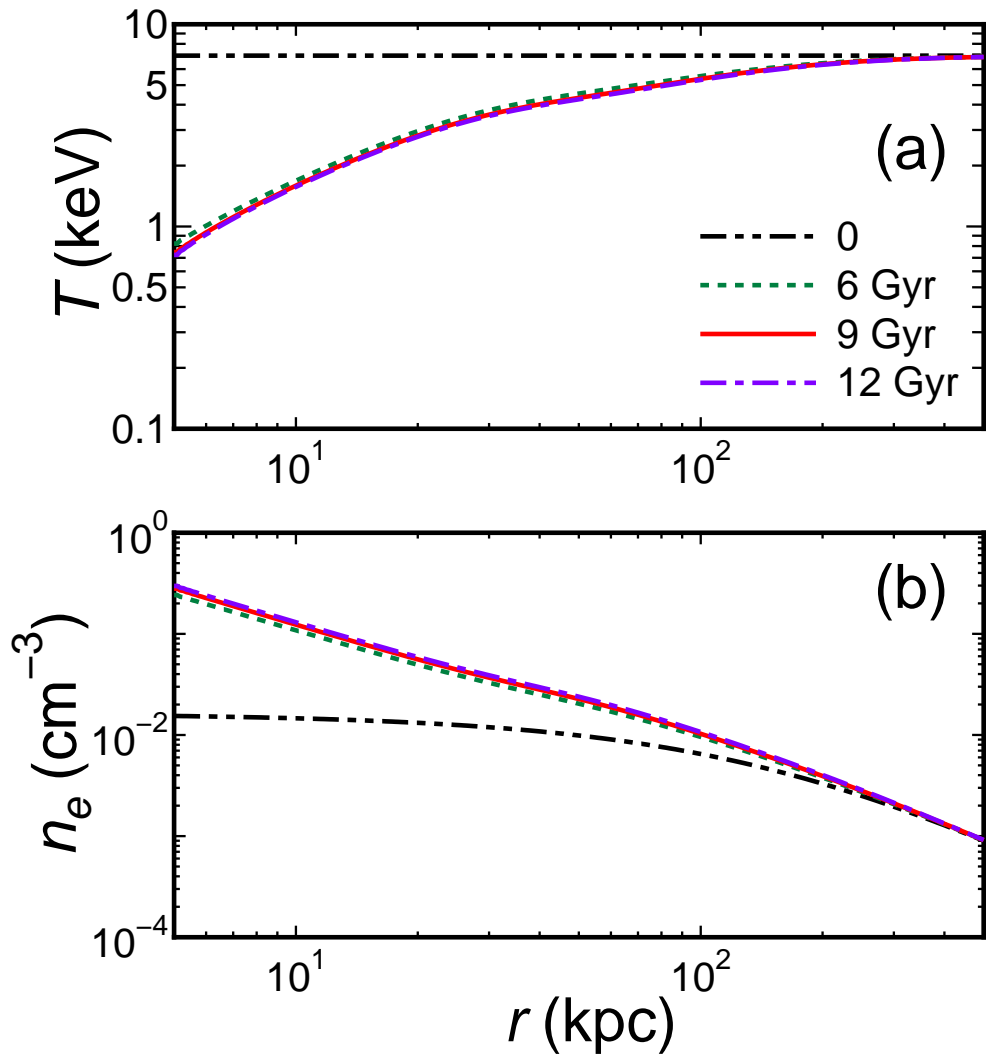} \caption{Same as Fig.~\ref{fig:Tn_lcr0}
but for the Model LCRs.} \label{fig:Tn_lcrs}
\end{figure}

\begin{figure}
\epsscale{.80} \plotone{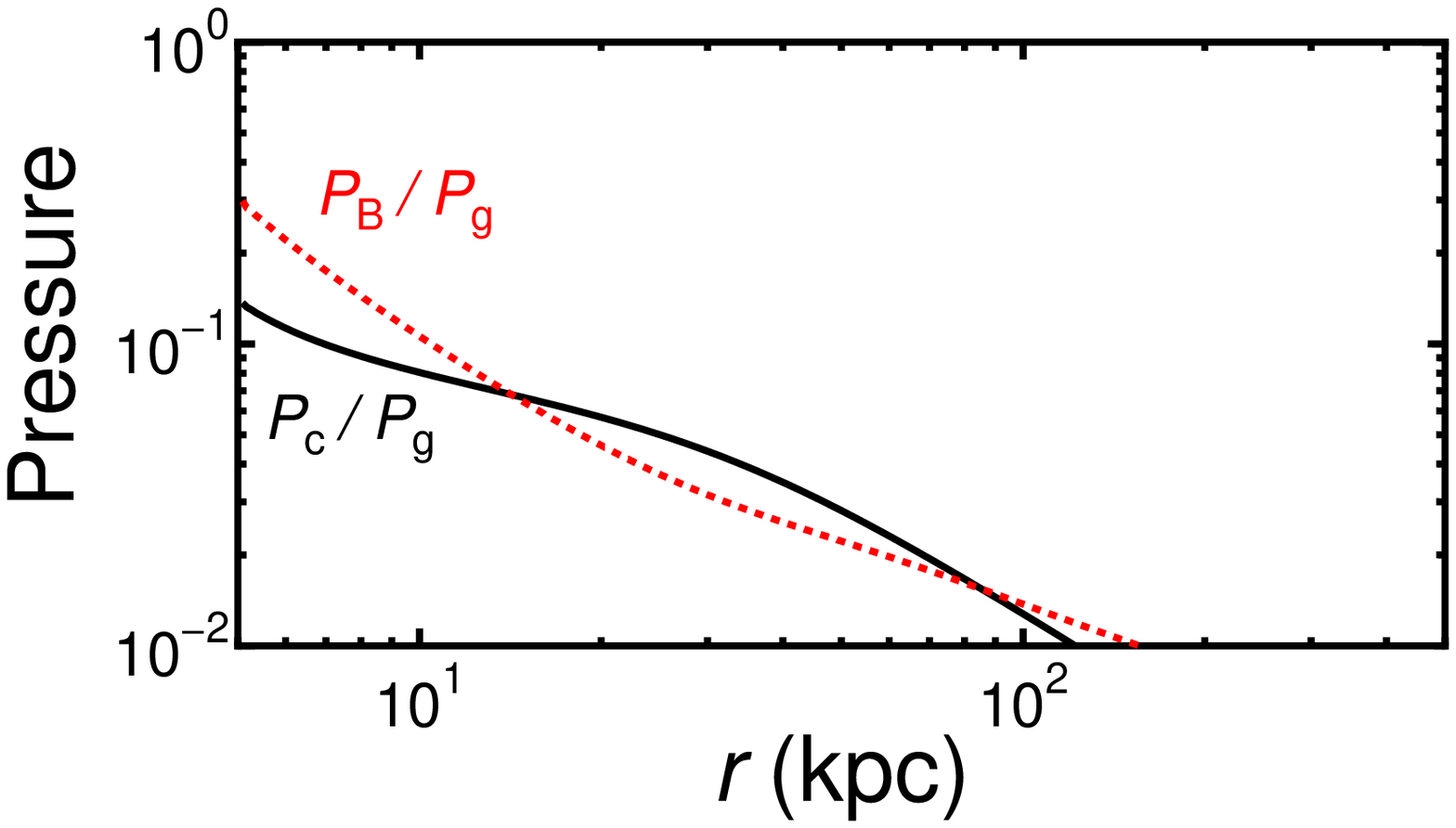} \caption{Same as
Fig.~\ref{fig:Pcb_lcr0} but for Model LCRs.}
\label{fig:Pcb_lcrs}
\end{figure}

\begin{figure}
\epsscale{.80} \plotone{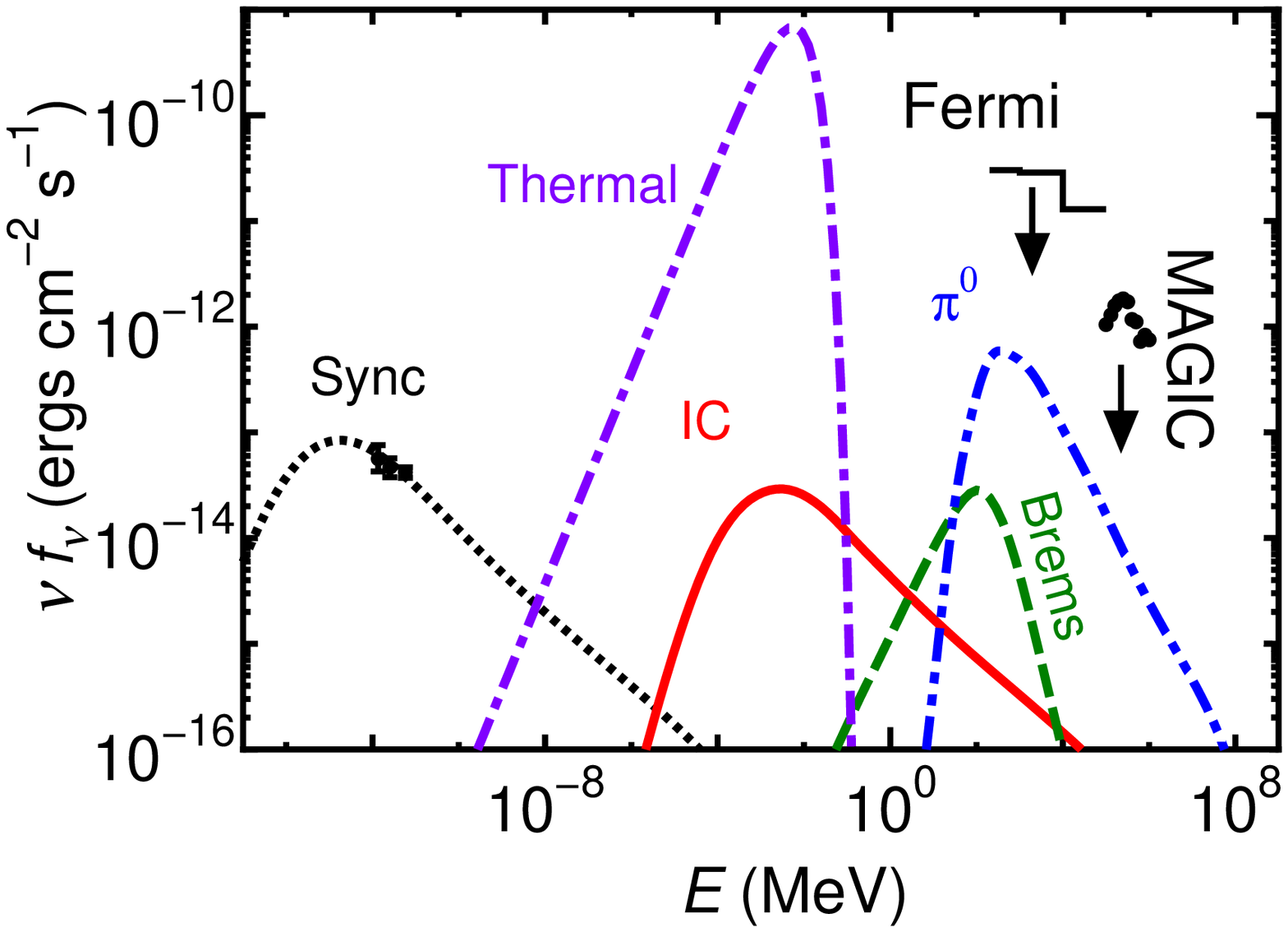} \caption{Same as Fig.~\ref{fig:lcr0}
but for Model LCRs and $M_{st}=2.4$.} \label{fig:lcrs}
\end{figure}

\begin{figure}
\epsscale{.80} \plotone{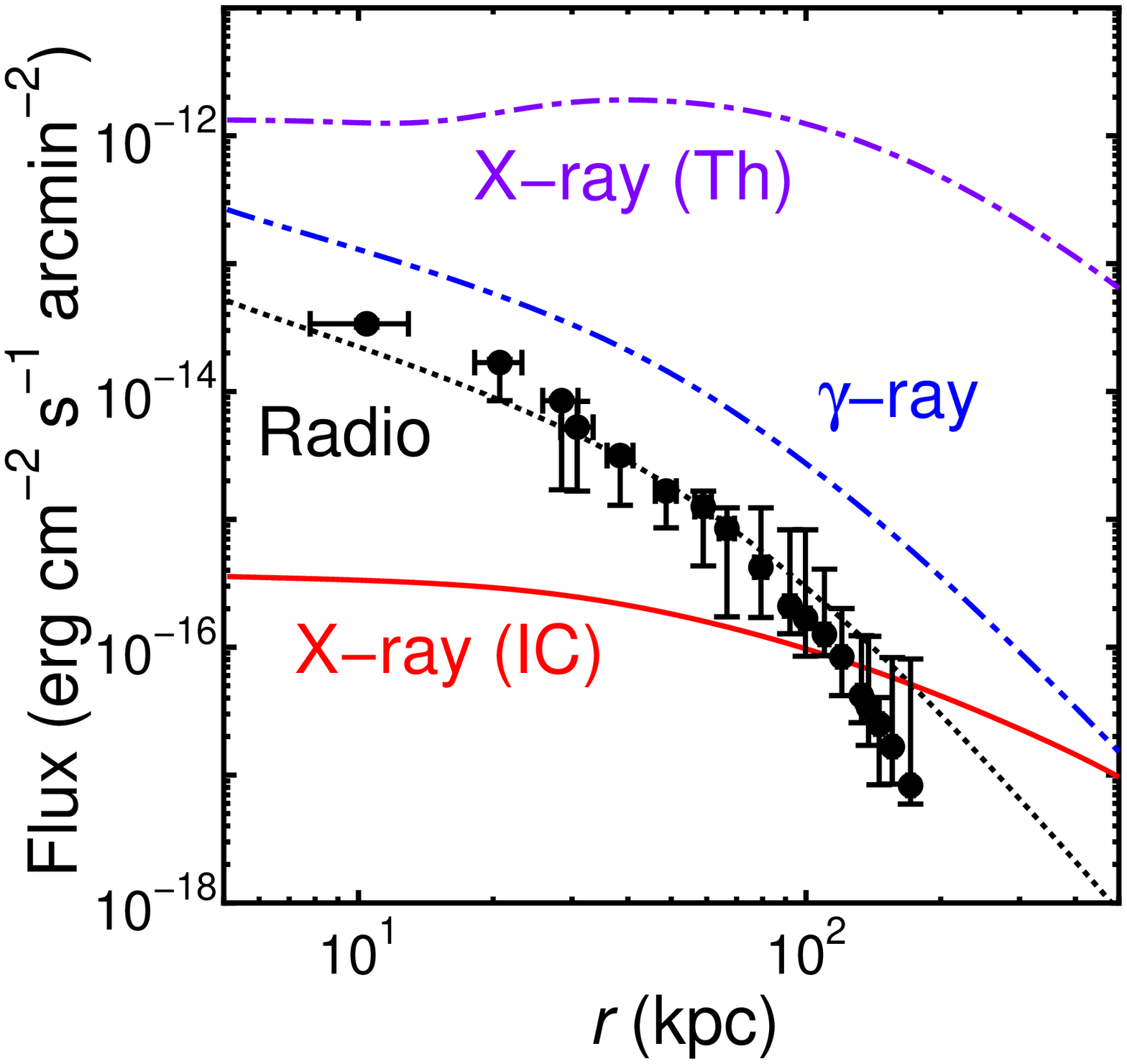} \caption{Same as
Fig.~\ref{fig:surf_lcr0} but for Model LCRs and $M_{st}=2.4$.}
\label{fig:surf_lcrs}
\end{figure}

Regardless of the streaming velocity, the CR spectra in the cores must
be steep, because if not, the luminosities are too large
(Figure~\ref{fig:lcr0_2}b); this is inconsistent with the small number
of clusters in which radio mini-halos have been observed \citep{gov09}
and the non-detection of gamma-rays from clusters. Because of the steep
spectra, future observations in the low-frequency radio band would be
useful. Thus, cool cores would be promising targets for radio telescopes
such as {\it LOFAR}. The number of mini-halos may increase as the
sensitivities of radio telescopes are improved.  In our model, we
assumed that CRs are mostly accelerated when the Mach number of the
shock is $M_s\sim M_{st}$. For real clusters, however, we expect that
the Mach number $M_s$ decreases and that the CR spectrum at the shock
steepens during the expansion of a cocoon. Thus, we expect that the
spectral index should increase outwards in the cluster, which has
actually been observed in the radio band \citep{sij93,git02}, although
the interferometric nature of these measurements might result in smaller
radio halos at higher frequencies (missing zero spacing problem). On the
other hand, observations in other bands would be difficult in the near
future (Figures~\ref{fig:lcr0} and~\ref{fig:scr0}). In the X-ray band,
IC emissions should be observed (Figures~\ref{fig:lcr0}
and~\ref{fig:scr0}). However, thermal emissions from cool cores are very
bright, which makes it difficult for the non-thermal emission to be
detected. For the Perseus cluster, \citet*{san05} claimed the detection
of non-thermal emission with a flux of $6.3\times 10^{-11}\rm\: erg\:
cm^{-2}\: s^{-1}$ between 2 and 10~keV. However, the detection was not
confirmed by later observations \citep{mol09,eck09}. Even with hard
X-ray telescopes that will be launched in the near future such as {\it
NuSTAR} and {\it Astro-H}, the detection may be difficult because of the
low surface brightness (Figures~\ref{fig:surf_lcr0}
and~\ref{fig:surf_scr0}). For the detection in the gamma-ray band, good
angular resolutions as well as sensitivities are required, because
gamma-rays could also be emitted from the central AGNs
\citep[e.g.][]{abd09,kat10}, which must be resolved.

The steep CR spectra mean that most of the CRs in cool cores have low
energies. Thus, indirect studies may be useful. For example, optical
filaments observed in cool cores may be heated by those CRs
\citep[see][]{fer09,bay10}. We note that the CR heating is locally
unstable, and that the filaments could be created through local thermal
instabilities (Paper~I). Moreover, our model does not require turbulence
for stable heating. Thus, cool cores in which turbulence is not
developing may be observed with detectors having high spectral
resolutions such as {\it Astro-H}, while the detection of turbulence
does not deny our model. Although we did not consider primary electrons,
they may be accelerated at shocks in cores in spite of the low Mach
numbers \citep{mat11}, and the emissions from them may be observed in
some clusters.

Finally, we caution the reader that we did not consider energy-dependent
diffusion of CRs, because we do not know the actual diffusion
coefficient in the ICM, especially away from the shock front
\citep[see][]{fuj11a}. If CRs with higher energies escape from the core
faster than those with lower energies, the energy spectrum could be
steep \citep*[e.g.][]{fuj09c,ohi10,ohi11}. Moreover, we did not include
the contribution of gamma-rays from CRs accelerated at cosmological
shocks and those from dark-matters \citep*[e.g.][]{tot04,pin10,pin11}.

\section{Conclusions}
\label{sec:conc}

We have investigated non-thermal emissions from cool cores heated by
CRs. For the distributions of CRs, we used the model in which the cores
are stably heated by CR streaming. CR protons interact with ICM protons
and produce secondary electrons and $\pi^0$-decay gamma-rays. We found
that the CR spectra must be steep in order to be consistent with
observations of a radio mini-halo. The steep spectra reflect the fact
that the CRs are accelerated at shocks with low Mach numbers ($\sim 2$)
in hot ICM. We have also studied the dependence on the CR streaming
velocity and found that the stronger shocks are required to be
consistent with the observations for the larger CR streaming
velocity. Since most of the CRs in cores have low energies, synchrotron
emissions from them should be observed in low-frequency radio
bands. Thus, the number of clusters that have radio mini-halos would
increase as the sensitivities of radio telescopes increase. On the other
hand, the detection in other bands such as the X-ray and gamma-ray bands
would be difficult in the near future. The low-energy CRs could be
studied by observing optical filaments that are often found in cool
cores.

\acknowledgments

We thank the referee for useful comments. This work was supported by
KAKENHI (Y.~F.: 23540308, Y.~O.: 21684014).

\clearpage

\end{document}